\documentclass{ws-jai}
\usepackage[flushleft]{threeparttable}
\usepackage{svg}
\usepackage{orcidlink}
\usepackage{graphicx}
\usepackage{subcaption}
\usepackage{booktabs}
\usepackage{siunitx}

\begin{document}

\markboth{Meghan A. Marangola}{Measuring Local Turbulence Along the Optical Path: 
Multi-beam Optical Seeing Sensor (MOSS)}

\title{Measuring Local Turbulence Along the Optical Path: \\
Multi-beam Optical Seeing Sensor (MOSS)}
\author{Meghan A. Marangola$^{1}$\orcidlink{0009-0003-2965-6695}, Elana K. Urbach$^{1}$\orcidlink{0000-0002-3205-2484}  and Christopher W. Stubbs$^{2}$\orcidlink{0000-0003-0347-1724}}

\address{$^{2}$Department of Physics, Harvard University, Cambridge, MA 02138, USA, meghanmarangola@college.harvard.edu\\
$^{2}$Department of Physics and Department of Astronomy, Harvard University, Cambridge, MA 02138, USA \\
}

\maketitle

\begin{abstract}
Deflection of light along the optical path is a major source of image degradation for ground-based telescopes. Methods have been developed to measure upper atmospheric seeing based on models of the turbulence in the atmosphere, but due to boundary conditions, transmission within telescope enclosures is more complex. The Multi-beam Optical Seeing Sensor (MOSS) directly measures  the component of the image quality degradation from inhomogeneity of the index of refraction within the telescope dome. MOSS outputs four near-parallel beams of light that travel along the optical path and are imaged by the telescope’s detector, landing like starlight on the telescope’s focal plane. By using a strobed light source, we can `freeze' the instantaneous index variations transverse to the optical path. This system captures both `dome' and `mirror' seeing. Through plotting the standard deviation of differential motion between pairs of beams, MOSS enables characterization of the length scale of turbulence within the dome. The temporal coherence of temperature gradients can be probed with different pulse lengths, and the spatial coherence by comparing pairs at different separations across the aperture of the telescope. Optical path turbulence measurements, alongside other telemetry metrics, will guide thermal and airflow management to optimize image quality. A MOSS prototype was installed in the 1.2 meter Auxiliary Telescope (AuxTel) at the Vera C. Rubin Observatory in Chile, and preliminary data constrain the optical path turbulence with a lower bound of 1.4 arcseconds. The optical path turbulence varied throughout the night of observing.
\end{abstract}


\section{Introduction}

Index variation within a telescope enclosure is among the sources of degradation that prevent ground-based telescopes from achieving the diffraction limit for point sources. In the era of survey science with the Vera C. Rubin Observatory \citep{ivezic2019}, all improvements to an instrument's resolution and reduction of distortion of celestial objects work toward realizing the instruments' full scientific capabilities.

\begin{figure}
    \centering
    \includegraphics[width=0.75\linewidth]{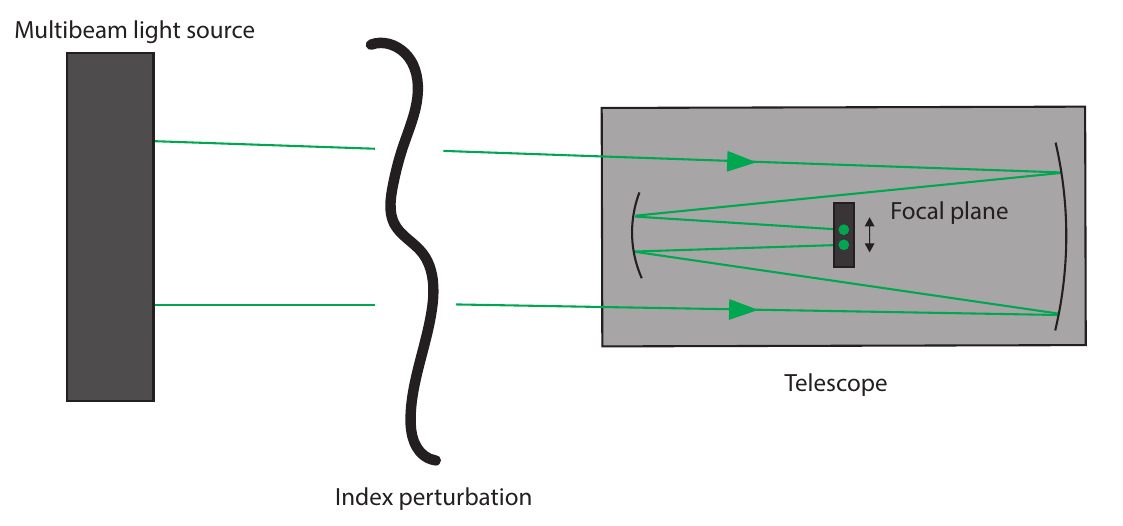}
    \caption{Conceptual illustration of a simplified MOSS-like device in operation. A source sends multiple light beams along the optical path through the telescope dome. The light encounters an index perturbation caused by temperature gradients and refracts. The angular shift from refraction alters the position of the light on the focal plane. Differential motion between PSF pairs at varying beam separations provides information about the characteristic length scale of the turbulent index perturbation the beams passed through. By using a strobed light source we can `freeze' the temporal index variations, with one flash per astronomical camera readout. 
}
    \label{fig:wavefront}
\end{figure}

Dome-seeing has previously been measured and modeled \citep{woolf1979, zago1997, racine1991}. Measurements of turbulence within telescope domes have also been tested \citep{kurmus2023, lai2019, bustostoko2018}. The Multi-beam Optical Seeing Sensor (MOSS) directly measures differential motion created by thermal variation along the optical path along which light from the stars travels \cite{marangola2025}. MOSS operates according to the conceptual design from \cite{stubbs2021} and depicted in Figure \ref{fig:wavefront}. Beams follow parallel trajectories and refract as they pass through turbulence in the dome, which changes their angle of arrival on the primary mirror and, thus, their position on the focal plane. By taking a series of images with a pair of beams, we can evaluate the differential image motion at that separation due to turbulence in the air. If the instrument has more than two beams, analyzing the correlation length of the differential motion yields the characteristic length scale of the turbulence in the dome.

The environment within the dome is much easier to control than the atmospheric conditions. Image quality optimization utilizing machine learning techniques has been proposed \cite{gilda2021}. Optimizing control systems using MOSS data will help mitigate the dome and mirror seeing by clearly separating the below-the-slit contributions to the FWHM.

\section{Methods} \label{sec:methods}

The MOSS system that we installed on the Auxiliary Telescope to the Rubin Observatory (AuxTel) dome is a proof of concept for the next-generation instrument to be installed on the Rubin Observatory's Simonyi Survey Telescope. The system sends four slightly off-parallel laser beams to the primary mirror of AuxTel, and when aligned, these beams continue to the focal plane. By plotting differential motion between PSF pairs against the spatial separation of the beams, we can characterize the dome and mirror seeing, or optical path turbulence along the telescope boresight.

\begin{figure}[htbp]
  \centering
  \begin{subfigure}{0.55\textwidth}
    \centering
    \includegraphics[width=\textwidth]{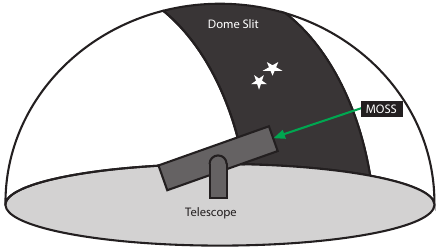}
    \caption{The MOSS system mounted within a telescope dome. The green arrow represents the path of MOSS beams to the telescope.}
    \label{fig:overview}
  \end{subfigure}
  \hfill
  \begin{subfigure}{0.4\textwidth}
    \centering
    \includegraphics[width=\textwidth]{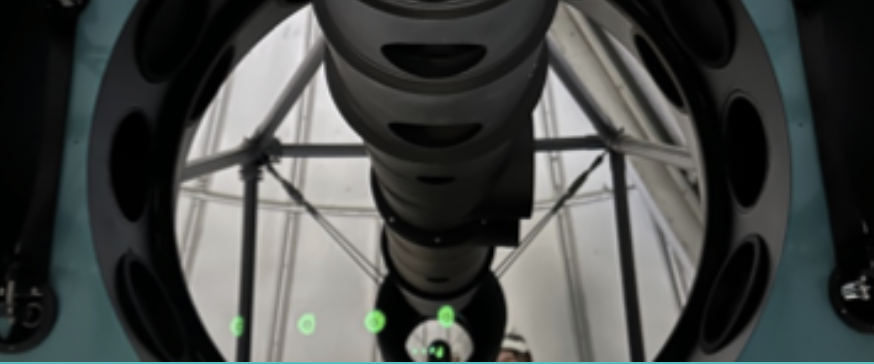}
    \caption{MOSS beams land on a chord of the primary mirror of the telescope. }
    \label{fig:chord}
  \end{subfigure}
  \caption{MOSS position within the telescope dome and path of the beams onto the pupil of the telescope. The telescope will alternate between pointing out through the slit and shifting to the side to point at MOSS. The alternating observing strategy can yield optical path turbulence comparisons for on-sky data throughout the night. }
  \label{fig:main}
\end{figure}

The turbulent layer within a telescope dome varies with wind speed, temperature, wind direction, and slit direction. The optical path turbulence thus evolves throughout each night, and monitoring over time is essential to understand the impact on the data. The MOSS system is designed for mounting on a telescope dome, as shown in Figure \ref{fig:overview}. The MOSS beams travel into the telescope, landing on the primary mirror as depicted in Figure \ref{fig:chord}. During observation, the telescope can quickly pan from taking on-sky data to point at the MOSS system. A series of 50 images takes less than five minutes, and data analysis would provide an optical path turbulence estimate for that time interval. By interleaving MOSS data with science observations, we can track the progression of the optical path turbulence during a night. This allows for more consistent correction procedures that adjust with feedback from daytime conditions.

On AuxTel, MOSS is mounted at approximately 25\textdegree~in elevation. We chose to mount the system adjacent to the dome slit for smaller average telescope re-pointing to take data. We incorporated remote adjustment into the instrument design; all beam pointing can be controlled remotely with an Intel NUC, using tip-tilt stages to adjust the orientation of the beamsplitters and mirrors.

\subsection{Components}

\begin{figure}[h]
    \centering
    \includegraphics[scale=1.5]{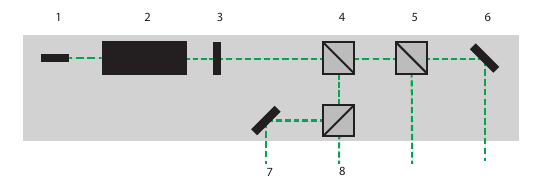}
    \caption{Schematic representation of the MOSS instrument, viewed from above.  A laser diode (1) outputs light pulses, which travel through a beam expander (2) so that their PSFs are smaller on the focal plane. The system then has an ND filter (3) on a flip mount for adjustability of the number of photons given a pulse length. A series of beamsplitters (4, 5, 8) and mirrors (6, 7) divide the light into four beams with roughly equal flux and are mounted on tip/tilt stages to allow for remote beam pointing adjustment. Beam 1 comes from mirror 7, Beam 2 from beamsplitter 8, Beam 3 from beamsplitter 5, and Beam 4 from mirror 6 corresponding to the labels in Figure \ref{fig:chord}. 
}
    \label{fig:conceptual diagram}
\end{figure}

The layout of the MOSS system components is depicted in Figure \ref{fig:conceptual diagram}. The MOSS light source is a Quarton VLM-520-61 LPO low-divergence laser diode, which has a wavelength of 515\ensuremath{\sim}530 nm and power output of less than 0.39 mW. As parallel light beams that hit the telescope's primary mirror converge to the same point on the focal plane, collimation is essential for resolving light sources; we chose a laser with a half angle divergence of less than 0.4 mRad. The laser diode is powered and controlled by a LabJack. 

The beam size on the focal plane was adjusted by sending the light through connected Thorlabs BE20-532 and BE05-532 beam expanders. This reduces the beam divergence in proportion to the expansion in beam size. The $\sim$ 50 mm diameter beams from MOSS yield diffraction-limited PSFs of around 3.5 arcsecond FWHM on AuxTel's 6.7 square arcminute field of view focal plane. This allows us, with adequate SNR, to obtain good centroid resolution. 

After the beam expander, an ND filter of OD 2.5 on a Thorlabs MFF102 flip mount attenuates the beam. The flip mount is primarily used during system alignment. The beam then passes through a series of Thorlabs BS031 beamsplitter cubes and Thorlabs BB2-E02 mirrors mounted on Thorlabs APY002 tip/tilt stages. The original micrometers on the stages were replaced with Thorlabs PIA13 piezo actuators. All components are mounted on a Thorlabs MB648 breadboard, and the beams were mounted 15.24 cm apart. The system was connected to the dome by mounting another MB648 breadboard to horizontal dome structures with hose clamps, then connecting the main MOSS board with a right angle bracket in the center and turnbuckles supporting the ends.

Motorized piezo actuators provide remote tip/tilt alignment capabilities and $\pm 4$\textdegree~of adjustment after the system is mounted. The layout of the beamsplitters and mirrors creates four beams of roughly equal flux. Beamsplitters with different reflection and transmission percentages could aid in scaling this system to more equal-flux beams. We aligned the beams to travel along the optical path, so they hit the telescope along a chord of the primary mirror. The successor system, now under development, uses a single reflection per beam to better match their intensities. 

Through LabJack control, the laser can emit pulses as short as 10 $\mu$s. We tested various pulse lengths and exposure times paired with different ND filters. Long 100 ms pulses averaged over the turbulence we wanted to freeze, and this suppressed the observed differential image motion. Very short 10 $\mu$s pulses were, we surmise, subject to thermal transient effects from the diode's nonuniform turn-on. We found that the optimal pulse for freezing the turbulent layer without adding laser turn-on effects was 0.1 ms, and an ND filter pairing of OD 2.5 gave the desired number of photons. Users can change the flip mount position to switch the strength of the ND filter, and remote adjustment to OD 3.5 allows 1 ms pulses if desired.

An Intel NUC operates the system's LabJack, piezo acutators, and flip mount. Users can remote into the computer, which is mounted on the side of the dome with the system. Power and ethernet are provided by a box that rotates with the dome.

\section{Analysis}

\begin{figure}[htbp]
  \centering
  \begin{subcaptionbox}{MOSS beams as imaged on the AuxTel focal plane. AuxTel has a 4k by 4k focal plane with 16 amplifiers. The non-gaussianity in beam shape comes from use of the laser diode.\label{fig:raw}}[0.47\linewidth]
    {\includegraphics[width=\linewidth]{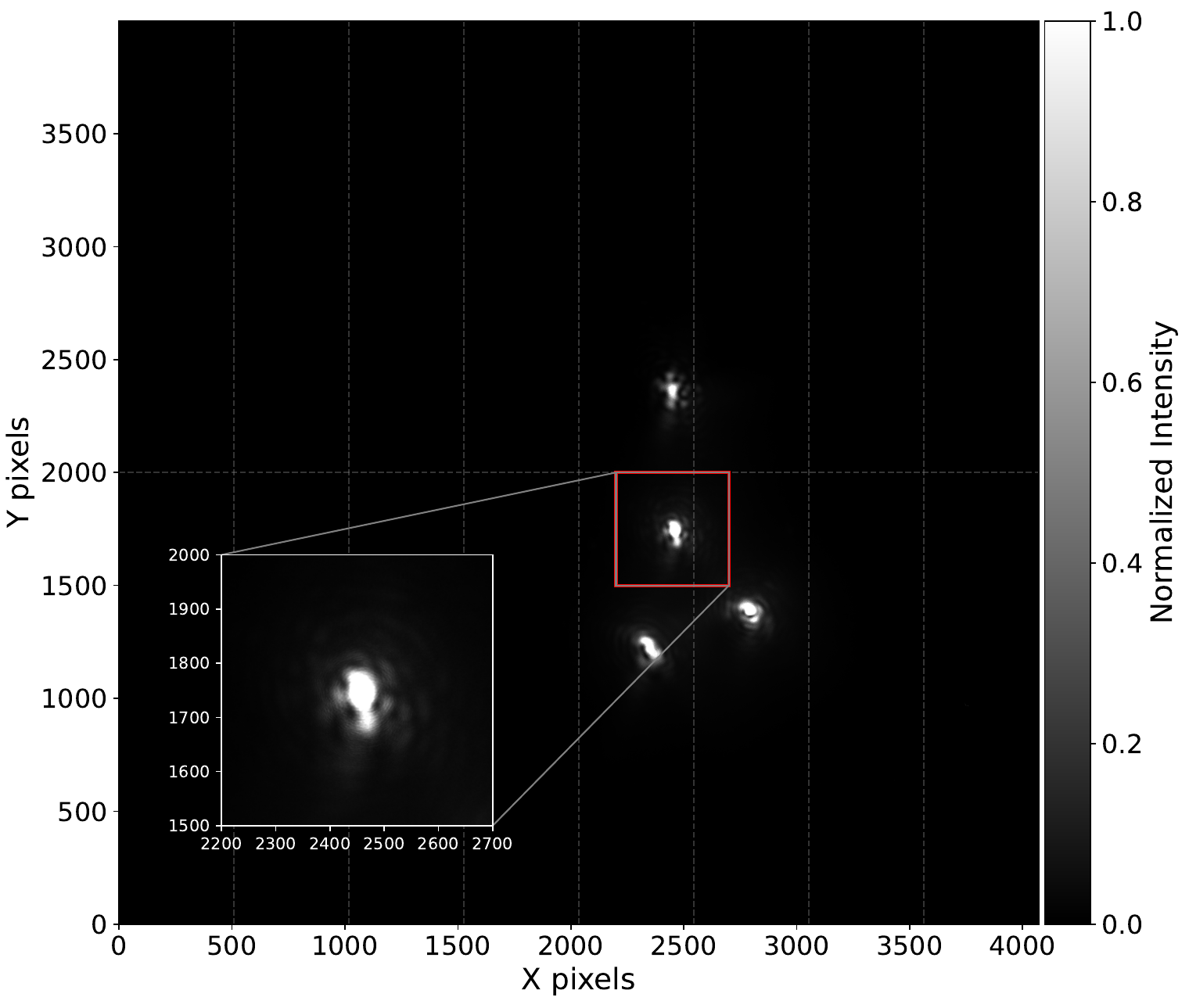}}
  \end{subcaptionbox}
  \hspace{1em}
  \begin{subcaptionbox}{Identical exposure to Fig. \ref{fig:raw} after convolution of the beam region with a Gaussian filter. Locations of centroids fit by the LSST Science Pipelines are plotted atop the image. \label{fig:centroids}}[0.47\linewidth]
    {\includegraphics[width=\linewidth]{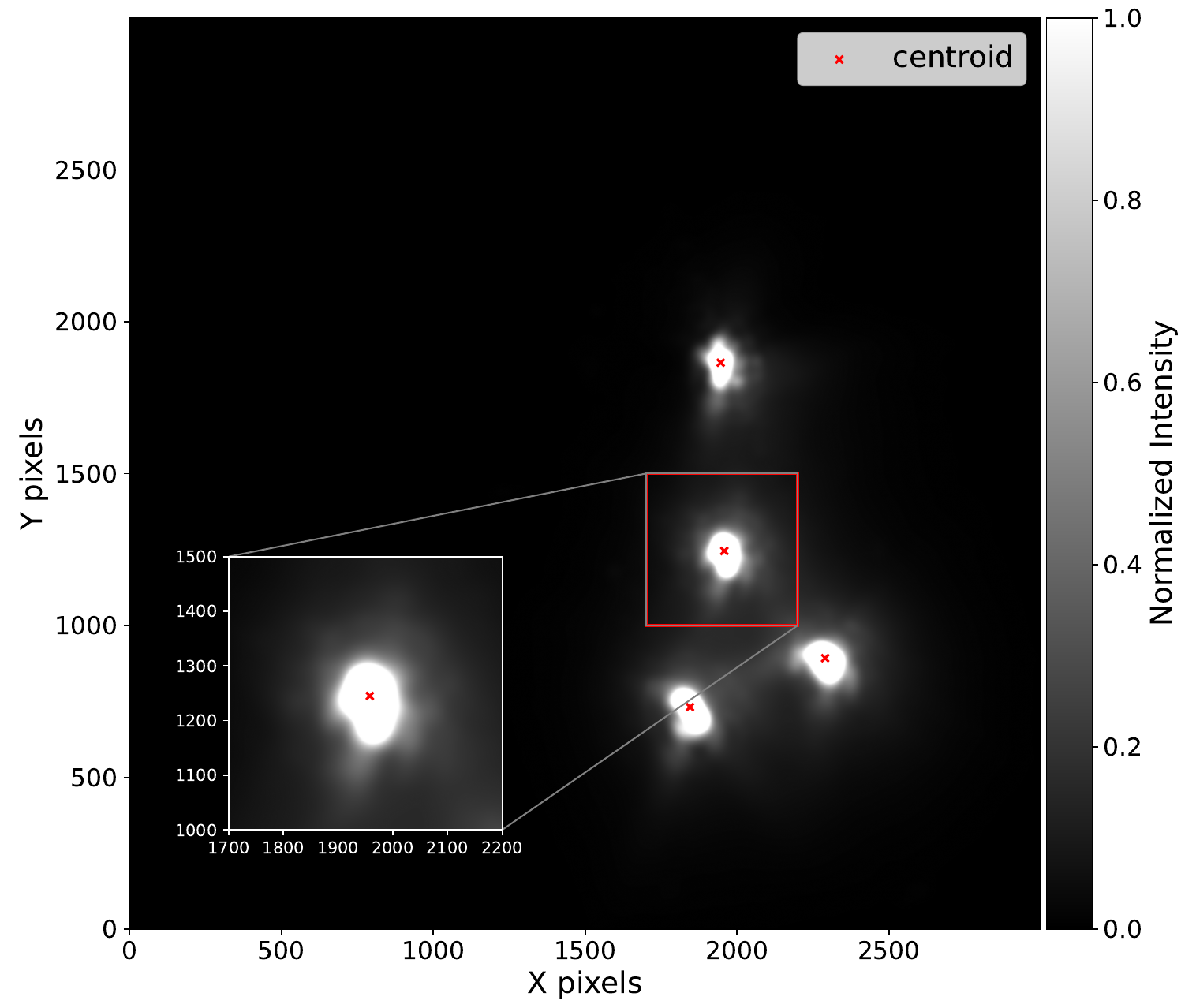}}
  \end{subcaptionbox}
  \caption{Comparison of a raw AuxTel image and the same image after convolution with a Gaussian kernel with FWHM of 35.32 pixels and centroid position calculation. The distance between the pairs of beams is calculated using these centroid locations. 
  }
  \label{fig:side_by_side}
\end{figure}

After installing the instrument in AuxTel, we performed engineering tests from 2024-05-12 at 12:20 (UTC) to 2024-05-13 at 06:10 (UTC), and from 2024-05-13 at 16:28 (UTC) to 2024-05-14 at 04:18 (UTC). We then took data with the instrument on the final night from 2024-05-14 at 14:01 (UTC) to 2024-05-15 at 05:20 (UTC). We took images of the MOSS beams using the camera on AuxTel with two second exposure times. The AuxTel camera comprises a 4k x 4k deep-depletion CCD, with 16 readout amplifiers. The camera readout time is two seconds. Each set of MOSS data must include enough images for the standard deviation of the differential motion to be statistically significant. The main set of data referred to in this paper was a sequence of 1000 frames, 878 of which passed centroiding quality cuts.

Figure \ref{fig:side_by_side} shows the data before and after processing; \ref{fig:raw} displays the raw PSFs on the focal plane. The imperfect Gaussian spatial profiles result from the laser diode. We convolve the raw image with a Gaussian kernel with a FWHM of 35.32 pixels; the same image as Fig. \ref{fig:raw} after this convolution is shown in Fig. \ref{fig:centroids}. The centroids are calculated with the LSST science pipelines. The average centroiding uncertainty was 0.00085 arcsec, which was computed for each PSF with values shown in Table \ref{table} by dividing its FWHM by its signal to noise ratio. 

\begin{table}
\centering
\caption{Beam Parameters}\label{table}
\renewcommand{\arraystretch}{1.3}
\begin{tabular}{|c|>{\centering\arraybackslash}p{3.5cm}|>{\centering\arraybackslash}p{4.0cm}|>{\centering\arraybackslash}p{4.5cm}|}
\hline
\textbf{Beam} & \textbf{Typical total flux (electrons)} & \textbf{Post-convolution FWHM (arcseconds)} & \boldmath$\sigma_\text{centroid} = \text{FWHM}/\sqrt{\text{Flux}} $ \textbf{(arcseconds)} \\
\hline
1 & \num{65e6} & \num{6.59} & \num{0.00082} \\
2 & \num{82e6} & \num{5.89} & \num{0.00065} \\
3 & \num{71e6} & \num{5.47} & \num{0.00065} \\
4 & \num{47e6} & \num{8.86} & \num{0.00130} \\
\hline
\end{tabular}
\end{table}

\begin{figure}[htbp]
    \centering
    \includegraphics[width=0.96\linewidth]{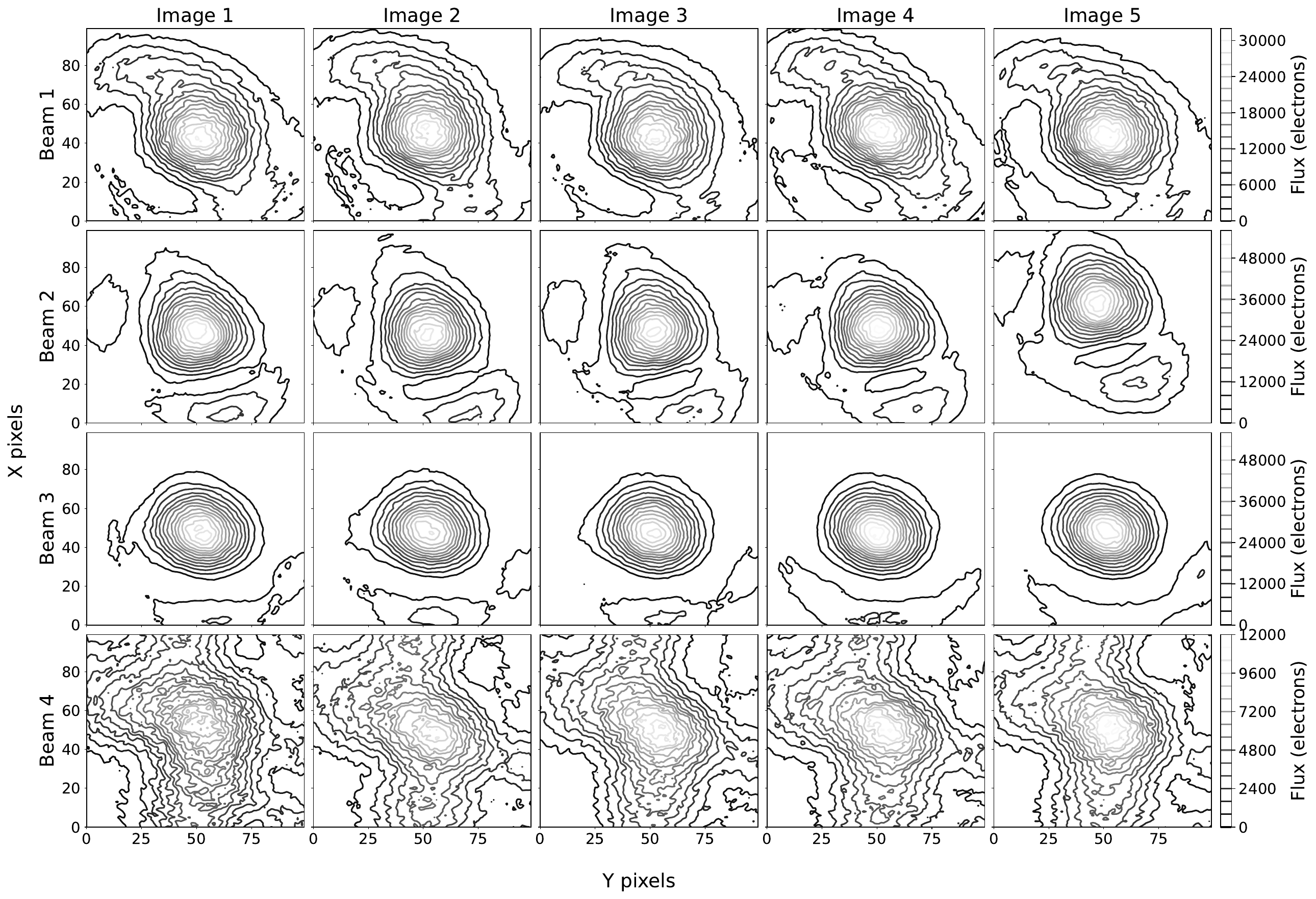}
    \caption{Contour plots of the four MOSS beams for a series of five consecutive images from the sequence of 878 analyzed in this paper. Images had a two-second exposure time and were separated by the two-second camera readout. The rows and columns correspond to beams and images, respectively. The PSFs vary between beams, but for a given beam, the PSF shape is stable across images. The consistent beam shapes prevent the asymmetry present in individual beams from impacting the differential motion analysis.}
    \label{fig:beam-contours}
\end{figure}

The centroid position is heavily influenced by beam shape. One potential error source would be centroid motion arising from the changing PSF shape of a single beam, which could be compounded if PSFs were highly variable across all beams. Figure \ref{fig:beam-contours} illustrates that the beam shape is stable for a given beam across a stack of consecutive images. The camera exposure time per image was two seconds, followed by the two-second camera readout. Although individual beams have distinct PSFs, the individual consistency of each beam across images enables reliable centroiding.

\begin{figure}[h]
    \centering
    \includegraphics[width=0.7\linewidth]{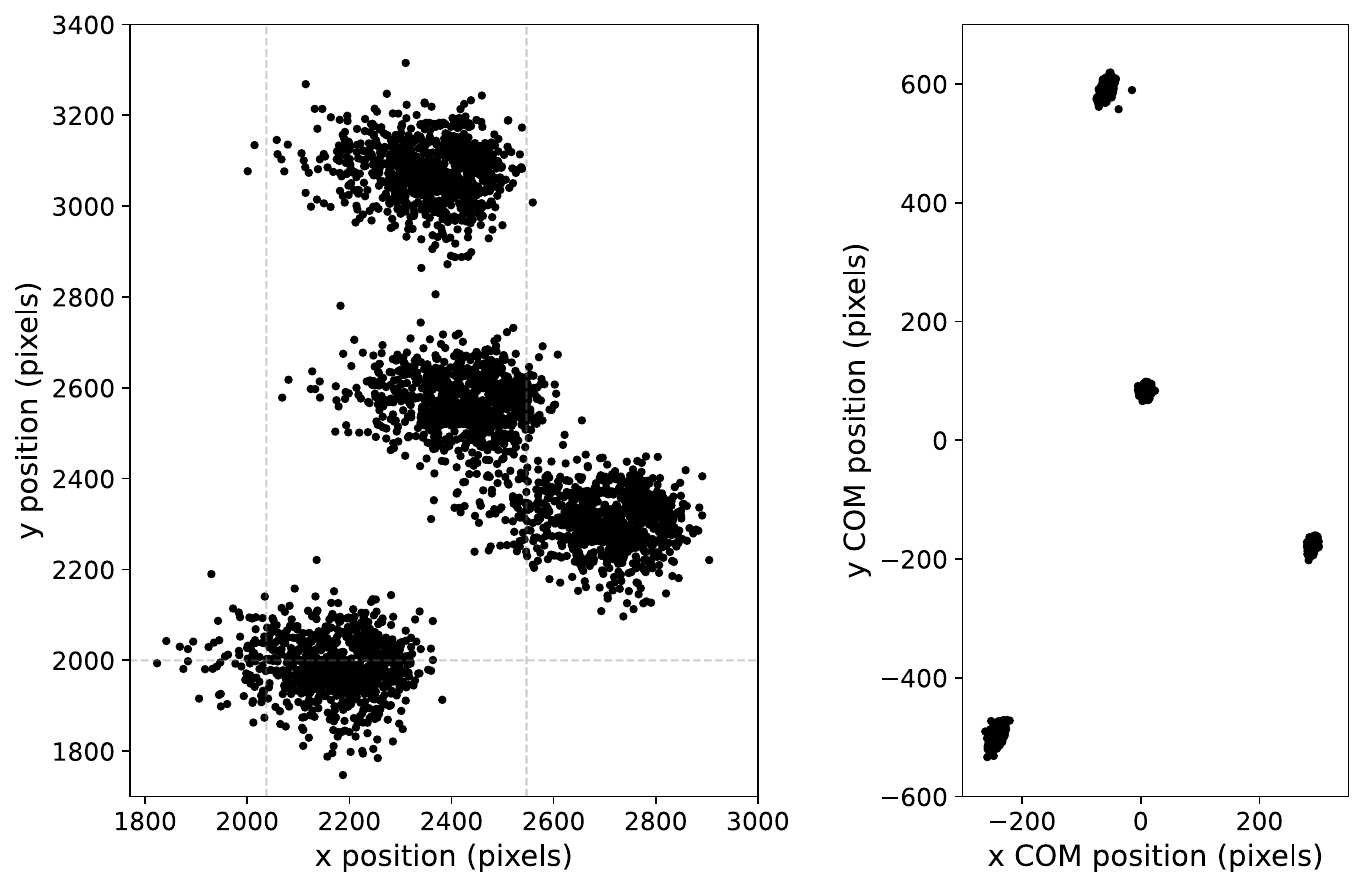}
    \caption{The panel on the left shows the fitted centroids over a sequence of 1000 images. The plot on the right shows the same information after subtracting out the position of the center of mass (COM) of the beams. The greatly reduced size of the clusters indicates that much of the motion of the beams is common mode. The subtraction between points to get the PSF separation is independent of the common-mode motion, as points are within the same image. When comparing between images, we separated the differential motion from the common-mode motion by changing the coordinates for each image to a COM frame.}
    \label{fig:COM}
\end{figure}

After finding the centroids for each beam in each image, we compare PSF separations between images. As we only care about differential values for this analysis, we ignore the common-mode motion of the beams. The center of mass moves significantly between images, as shown in Figure \ref{fig:COM}. These data were taken on a particularly windy night, so movement of the dome likely caused this motion. There could also have been slight telescope movements that changed the absolute position of all beams on the focal plane.

\begin{figure}
    \centering
    \includegraphics[width=0.7\linewidth]{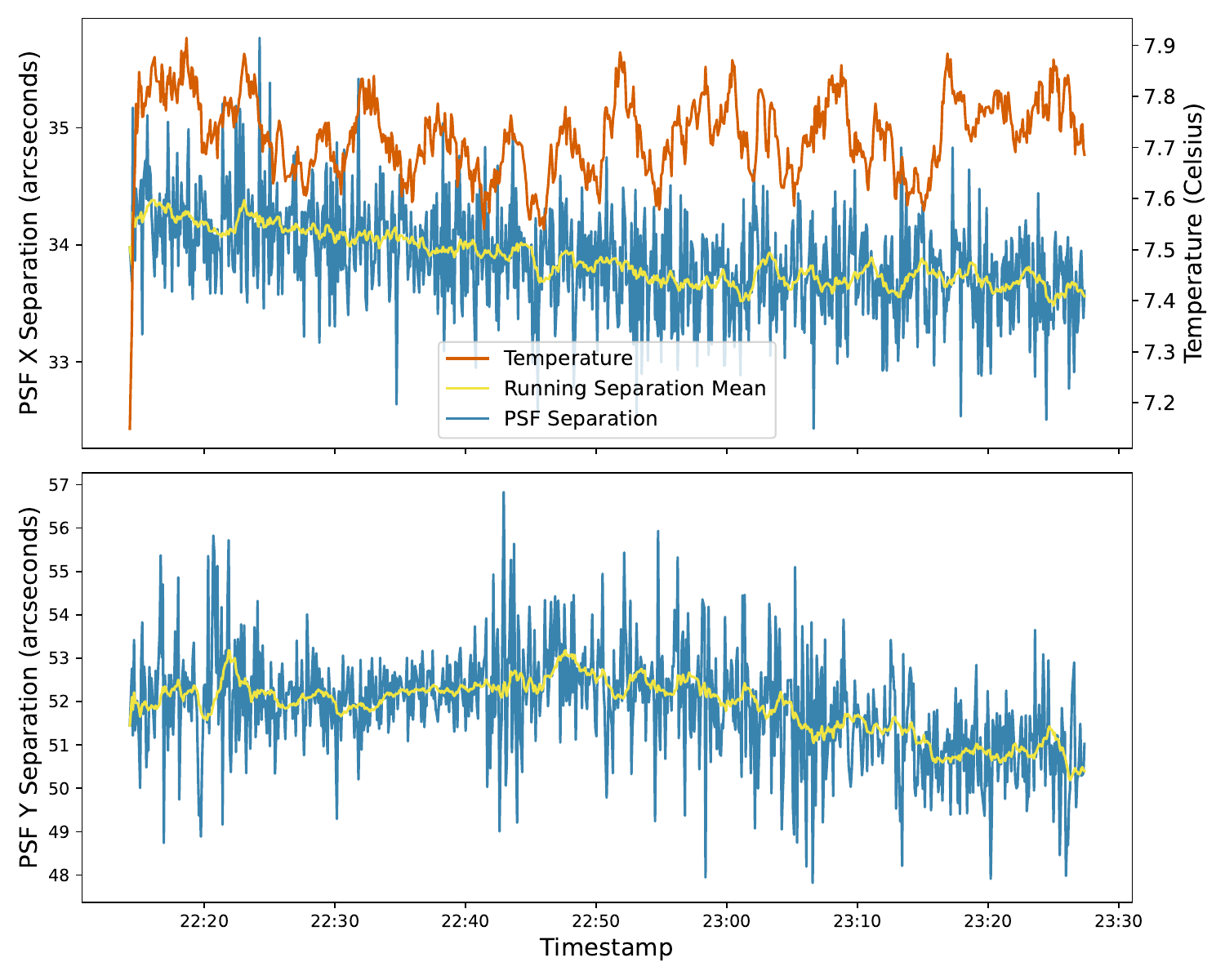}
    \caption{This graph displays PSF separation in X and Y for a single PSF pair over a series of images measured in time on the x-axis. The top plot shows the separation in X and the temperature over the same time period, while the bottom plot shows the separation in Y. The running means of the separations in X and Y with windows of 20 points are shown in yellow. There is no short-term correlation between separation and change in temperature. We subtracted off the running mean before calculating the standard deviations of separation for each image pair. }
    \label{fig:separationDrift}
\end{figure}

Once we had the PSF separation for each pair of beams, we checked to ensure that the standard deviation of PSF separation would not be dominated by shifts over time. Plotting PSF separation over a long sequence of images illustrates the oscillation of the separation value and the long-term drift of the PSF separation. The plot for a single pair is shown in Figure \ref{fig:separationDrift}, alongside the running mean in X and Y and the temperature. There is no significant correlation of beam separation with temperature, indicating no mechanically temperature dependent system component. We subtract off the timestamped running mean with data taken at that time so that our standard deviation measure probes the turbulence and not the drift of a single beam in the apparatus.

The PSF locations do not necessarily correlate with the separation between beams on the optical bench.  We identified the beams by blocking each and taking an image. The identification could also be accomplished by moving each beam a small amount between consecutive images. After identifying the beams, we calculate the separations between each PSF pair, for all six pairs. Then we take the standard deviation of that pairwise separation over the stack of images we are analyzing. This yields the points on the graph in Figure \ref{fig:1000}. The 67\% confidence interval error bars result from the standard deviations calculated from pairwise separations bootstrapped from the stack of images.

\section{Results} \label{sec:results}

\begin{figure}[h]
    \centering
    \includegraphics[width=0.7\linewidth]{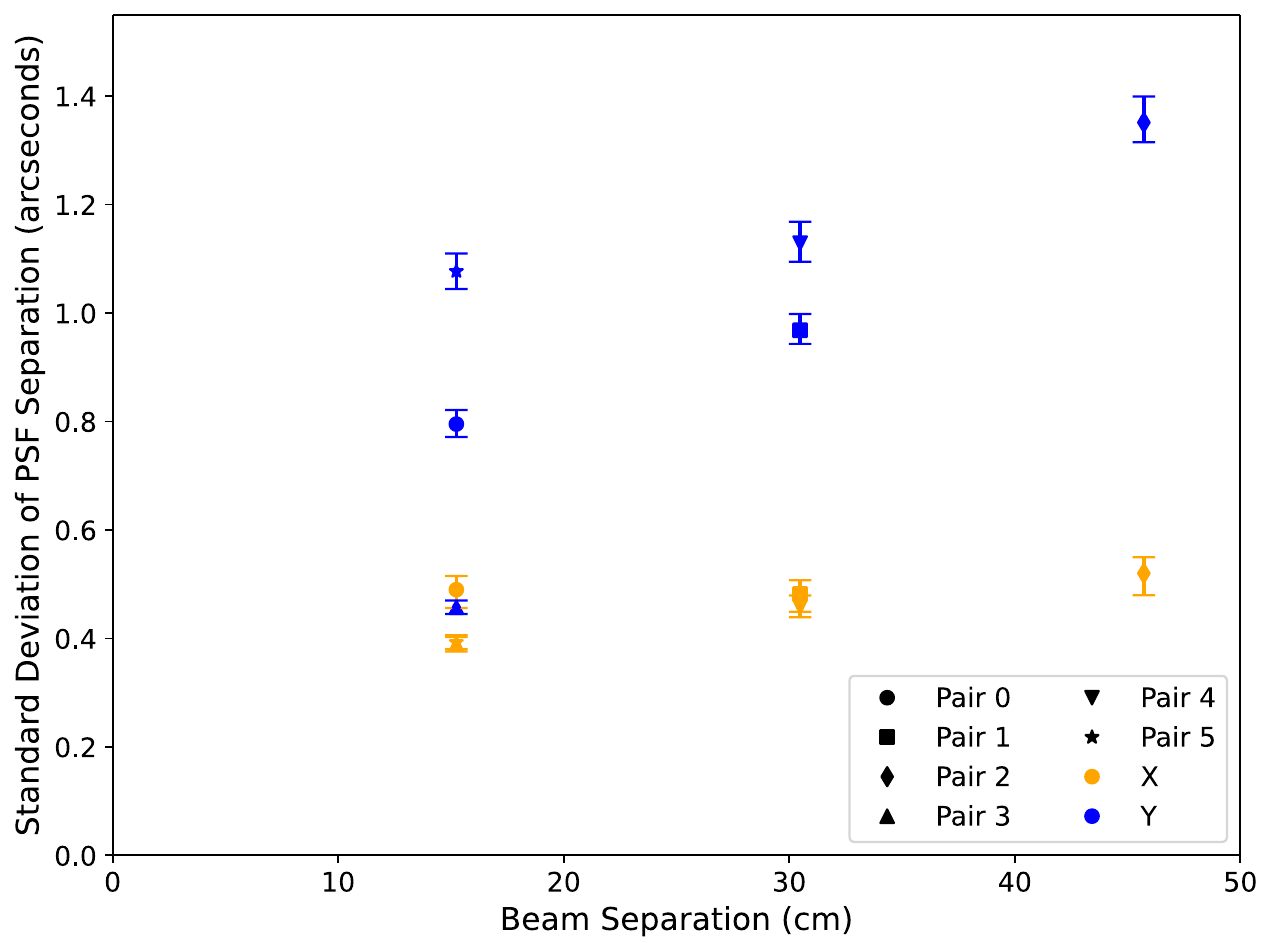}
    \caption{Optical path turbulence and 1\,$\sigma$ error bars for a sequence of 1000 images taken with the dome closed. The standard deviation of the PSF separation increases with their beam separation at the pupil. As the incoming beams mimic light from a single star that passes through a turbulent layer, taking the standard deviation of PSF separation measures maps to the diameter of the dome and mirror seeing disc. For each PSF pair, the seeing in Y is greater than the seeing in X, indicating greater vertical temperature fluctuations in the dome that lead to this positional variability. There is not yet the plateau in Y we would expect from completely uncorrelated beam motions, but the points across beam separation in X are flatter, suggesting a longer correlation length scale in Y. The in-dome contribution to seeing can be constrained with a lower bound of 1.4".}
    \label{fig:1000}
\end{figure}

We analyzed the standard deviation of PSF separations across 1000 images taken with the dome closed. The standard deviation of differential PSF motion is plotted against the beam separation in Figure \ref{fig:1000}. There is an increase in the standard deviation of PSF separation with the increase in beam separation, which we expect given that the beams are going through different patches of the turbulent layer. The standard deviations are much larger Y than in X, indicating a larger index variation vertically in the dome that causes more beam position variation in the Y direction. The standard deviation plateaus in X while in Y, the characteristic length scale of the turbulence has not been reached as the standard deviation continues to increase with beams farther apart. Thus we have a lower bound on optical path turbulence of 1.4 arcseconds. Even at the same spatial separation in the dome, different pairs have different standard deviations, so the turbulence pattern is not invariant under translation. When we plotted images in smaller subsets, we observed that the coherence time of the Y standard deviation points was shorter than the complete 1000-image set. Using the average centroiding uncertainty, we computed the uncertainty in the separation value to be 0.0012 arcseconds, which does not account for the measured differential motion.

We also explored the evolution of seeing over time with the dome open in Figure \ref{fig:evolution}. The night was particularly windy; average wind speeds increased and gusts blew up to 17 m/s. The values for optical path turbulence increased from the 1.4 arcsecond lower bound in Figure \ref{fig:1000}, and the trajectory between points flattened out, indicating that we reached the coherence length scale in this configuration. Temperature sensors in the dome during The X and Y values in these graphs show smaller separations because the wind blowing through the dome weakened the vertical temperature gradient.

\begin{rotatefigure}
    \centering
    \includegraphics[width=\textheight]{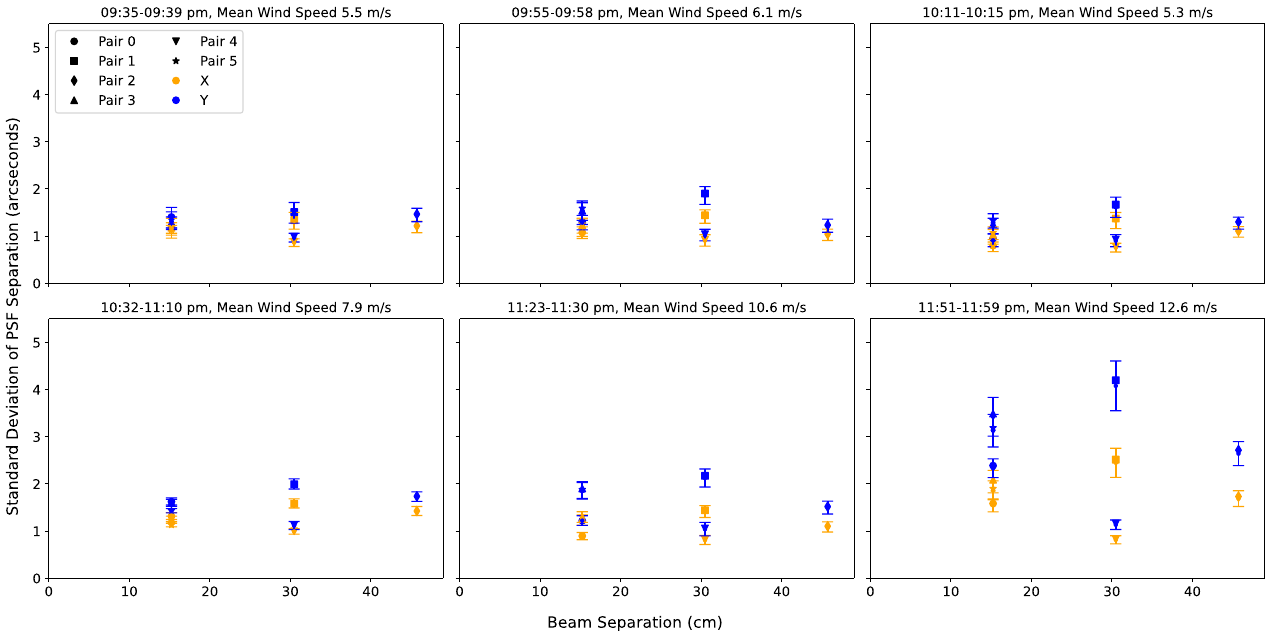}
    \caption{Optical path turbulence and 1\,$\sigma$ error bars for a series of image sequences taken with the telescope dome open. Perhaps the high wind speeds contributed to the increase in optical path turbulence values, which are higher than in Figure \ref{fig:1000}. Wind gusts exceeded 15 m/s later in the night, and we observed that the optical path turbulence increased significantly when wind speed spiked. The difference in seeing in the vertical and horizontal directions is no longer significant, and the graphs appear flatter, indicating that the characteristic length scale of the turbulence has been reached.}
    \label{fig:evolution}
\end{rotatefigure}

\section{Discussion and Future Work}
The MOSS instrument probes the effects of optical path turbulence along the exact path starlight travels. On AuxTel, MOSS revealed greater seeing and a longer correlation length in Y (vertical) than in X (horizontal), and an overall lower bound of 1.4" on optical path turbulence. Investigation must be done to determine the optimal number of MOSS images in a sequence and frequency of sequences during a night of observing. Future work will also include an exploration of temporal coherence of the optical path turbulence with differing pulse lengths. Additionally, joint analysis with the anemometer and thermal data in the Rubin enclosure should be conducted.

The AuxTel results are promising evidence to motivate installation of a similar system on other ground-based telescope domes. Future versions may find that increasing the number of beams and extending the length of the beam separation leads to better constraints on optical path turbulence, especially on larger telescopes like the Simonyi Survey Telescope.

\section*{Acknowledgments}
We are grateful for support from the United States Department of Energy under Cosmic Frontier award DE-SC0007881, from Schmidt Sciences, and from Harvard University. We thank Craig Lage, Freddy Muñoz Arancibia, Mario Rivera, Alysha Shugart, Ivan Gonzalez, and Hernán Stockebrand for their hands-on assistance on Cerro Pachón. Thank you to Kevin Fanning, Karla Aubel, and Yijung Kang for being our observing specialists during engineering tests. Thank you to Merlin Fisher-Levine and Leanne Guy for integrating our analysis with the LSST Science Pipelines.

\end{document}